\documentclass[showpacs,amsmath,amssymb,superscriptaddress,floatfix,reprint,aps,pra]{revtex4-1}

\usepackage{amsfonts}
\usepackage{mathrsfs}
\usepackage{amsmath}
\usepackage{graphicx}
\usepackage{grffile}
\usepackage{subfigure}

\newcommand{\ket}[1]{\mbox{$|#1\rangle$}}
\newcommand{\bra}[1]{\mbox{$\langle #1|$}}

\begin{document}

\title{Compact source of narrow-band counterpropagating polarization-entangled photon pairs using a single dual-periodically poled crystal}

\author{Yan-Xiao Gong}
\affiliation{Department of Physics, Southeast University, Nanjing,
211189, People's Republic of China} \affiliation{National Laboratory
of Solid State Microstructures and School of Physics, Nanjing
University, Nanjing, 210093, People's Republic of China}

\author{Zhen-Da Xie}
\affiliation{National Laboratory of Solid State Microstructures and
School of Physics, Nanjing University, Nanjing, 210093, People's
Republic of China}

\author{Ping Xu}
\email{pingxu520@nju.edu.cn} \affiliation{National Laboratory of
Solid State Microstructures and School of Physics, Nanjing
University, Nanjing, 210093, People's Republic of China}

\author{Xiao-Qiang Yu }
\affiliation{Department of Physics, Southeast University, Nanjing,
211189, People's Republic of China}

\author{Peng Xue}
\affiliation{Department of Physics, Southeast University, Nanjing,
211189, People's Republic of China}

\author{Shi-Ning Zhu}
\affiliation{National Laboratory of Solid State Microstructures and
School of Physics, Nanjing University, Nanjing, 210093, People's
Republic of China}

\begin{abstract}

We propose a scheme for the generation of counterpropagating
polarization-entangled photon pairs from a dual-periodically poled
crystal. Compared with the usual forward-wave type source, this
source, in the backward-wave way, has a much narrower bandwidth.
With a $2$-cm-long bulk crystal, the bandwidths of the example
sources are estimated to be $3.6$ GHz, and the spectral brightnesses
are more than $100$ pairs$/$(s GHz mW). Two concurrent
quasi-phase-matched spontaneous parametric down-conversion processes
in a single crystal enable our source to be compact and stable. This
scheme does not rely on any state projection and applies to both
degenerate and non-degenerate cases, facilitating applications of
the entangled photons.

\end{abstract}


\maketitle

\section{introduction}\label{sec-intro}

Polarization-entangled photons play a key role not only in testing
the foundations of quantum mechanics~\cite{review:nonlocality2005}
but also in various photonic quantum
technologies~\cite{review:obrien:techonlogies}. A compact, robust,
and high-brightness source of polarization-entangled photons is
therefore desirable for practical implementation of a variety of
entanglement-based applications.

Spontaneous parametric down-conversion (SPDC) in nonlinear crystals
is a successful technique to generate polarization-entangled photon
pairs. A typical method involves using the type-II birefringence
phase-matching (BPM) in a nonlinear
crystal~\cite{PDC:entanglement:Kwiat1995}, such as beta barium
borate (BBO). However, only a small fraction of the total emitted
photons, the intersecting locations of two non-overlapping cones,
are polarization-entangled, and therefore, such a source is
inefficient. A more efficient source consists of two type-I
nonlinear crystals via
BPM~\cite{PDC:entanglement:Kwiat1999,Source:twoI:Kim2001}, from
which polarization-entangled photons are emitted in a cone. However,
generally only a small fraction of the cone is collected for use,
and thus, such a source is again less efficient.

One way to solve the inefficiency problem in the conelike sources is
by means of quasi-phase-matching
(QPM)~\cite{QPM:Armstrong1962,QPM:Franken1963} in periodically poled
(PP) crystals~\cite{source:PPLN:tanzilli2001}, such as periodically
poled lithium niobate (PPLN) and periodically poled potassium
titanyl phosphate (PPKTP). QPM has advantages over BPM due to its
higher efficiency and the fact that it enables flexible
frequency-tunable processes. In particular, QPM enables the photon
pairs in a collinear and beam-like configuration. Consequently, it
is possible to make a much bigger fraction of the created photons
polarization-entangled than conelike sources, thus leading to more
efficient sources. However a new problem arises, namely the need to
spatially separate the collinear photon pairs.

A simple method to solve this problem is by using dichroic mirrors
when the photon pairs are generated at substantially different
frequencies. In this way, several non-degenerate polarization
entanglement sources have been designed by coherently combining two
SPDC sources at a polarizing beam
splitter~\cite{Source:PP:entangled:yoshizawa2004,Source:PP:entangled:Konig2005,Source:PP:entangled:Jiang2007,Source:PP:entangled:Lim2008b,*Source:PP:entangled:Lim2008,source:PP:entangled:Sauge2008,Source:PP:entangled:Hentschel2009,Source:PP:entangled:Fiorentino2008},
by manipulating  polarization
ququarts~\cite{Source:qudit:Moreva2006}, by overlapping two cascaded
PP
crystals~\cite{Source:PP:entangled:Pelton2004,Source:PP:entangled:Ljunggren2006}
, or by two cascaded \cite{source:PP:suhara} or
concurrent~\cite{Source:PP:entangled:Thyagarajan2009,source:PP:Levine2011}
SPDC processes in a single PP crystal. Theses non-degenerate sources
have various applications, for instance, in quantum
communication~\cite{review:communation}. However, in many
entanglement-based applications, for example, in quantum
computation~\cite{review:Kok2007}, frequency-degenerate
polarization-entangled photons are required. A straightforward way
to build degenerate entangled sources based on PP crystals is by
separating collinear orthogonally polarized photon pairs with a beam
splitter followed by twofold coincidence measurement as a
postselection~\cite{Source:PP:entangled:Kuklewicz2004}. However,
this method suffers a $50\%$ loss. A postselection-free method
employs interferometers to combine two pairs of orthogonally
polarized
photons~\cite{Source:PP:entangled:Fiorentino2004,*Source:PP:entangled:Kim2006,*source:PP:entangled:Kuzucu2008,Source:PP:entangled:Fedrizzi2007},
but such interferometric sources (also the non-degenerate sources in
Refs.~\cite{Source:PP:entangled:yoshizawa2004,Source:PP:entangled:Konig2005,Source:PP:entangled:Jiang2007,Source:PP:entangled:Lim2008b,*Source:PP:entangled:Lim2008,source:PP:entangled:Sauge2008,Source:PP:entangled:Hentschel2009,Source:PP:entangled:Fiorentino2008})
require stringent phase control and stabilization.

Another problem of SPDC sources lies in the broad bandwidth
determined by the phase-matching condition, which is usually on the
order of several THz or hundreds of GHz. The broadband SPDC source
becomes very dim in many applications requiring narrow-band photons,
such as long-distance fiber optical quantum communication
($\sim$GHz~\cite{source:narrow:Halder2005}), strong interaction of
the photons with atoms and molecules($\sim$MHz~\cite{repeater:2001},
and recently relaxed to several
GHz~\cite{memory:gisin2008,memory:Reim2010}), and interference of
independent sources without time synchronization
($\sim$GHz~\cite{source:PPLN:Halder2007}). Passive filtering is a
straightforward way to obtain narrow-band
sources~\cite{source:narrow:Halder2005,source:PPLN:Halder2007}, but
it will greatly reduce the generation rate.  Cavity-enhanced SPDC
can provide high-brightness narrow-band photon
paris~\cite{Source:narrow:ou1999,source:PP:bao2008,source:pp:narrow:Scholz2009}.
However, additional spectral filtering is required to obtain
single-mode output due to the broad gain bandwidth.

In this paper, we succeed in solving all the above problems by
building a compact and narrow-band polarization entanglement source
based on the backward-wave type SPDC in a dual-periodically poled
crystal. The backward-wave type
SPDC~\cite{Source:PP:backward:Christ2009,Source:PP:backward:Suhara2010,Source:PP:backward:Chuu2011},
has a much narrower bandwidth than the forward-wave interaction. The
counterpropagating photon pair generation has also been extensively
studied in waveguide
structures~\cite{source:SPDC:counter:DeRossi2002,source:SPDC:counter:Booth2002,*source:SPDC:counter:Walton2003,*source:SPDC:counter:walton2004,source:SPDC:counter:Jr2008,source:SPDC:counter:Ravaro2005,*source:SPDC:counter:Lanco2006,*source:SPDC:counter:Orieux2011}.
Moreover, it not only has the same advantage as the usual collinear,
beam-like output SPDC on photon collection and overlapping for
possible polarization entanglement, but it also does not suffer from
the problem of spatial separation. Our scheme relies on the
coherence of two concurrent backward-wave type SPDC processes in a
single PP crystal, rather than any interferometer and postselection.
Furthermore, this scheme can work in frequency degenerate and
non-degenerate cases, for which we design two experimentally
feasible structures, respectively. With a $2$-cm-long bulk crystal,
the bandwidths of the two sources are estimated to be $3.6$ GHz,
with spectral brightnesses of $115$ and $154$ pairs$/$(s GHz mW),
respectively.

The rest of this paper is organized as follows. In the next section,
we give a description of the dual-periodically poled crystal and
design the structures required in our scheme. In
Sec.~\ref{sec-SPDC}, we introduce our scheme and make detailed
calculations on the sources we propose. Sec.~\ref{sec-con} are
conclusions.

\section{Description of a dual-periodically poled crystal}\label{sec-crystal}

QPM originates from modulation of the second-order nonlinear
susceptibility $\chi^{(2)}$. It has been advanced to a variety of
domain structures which allow multiple and flexible nonlinear
processes in a single crystal, leading to compact and integrated
devices. A dual-periodic structure is one of the QPM structures,
which permits two coupled optical parametric
interactions~\cite{QPM:dual:Chou1999,QPM:dual:Liu2002}. Here, taking
the potassium titanyl phosphate (KTP) crystal for example, we design
a dual-periodic structure to satisfy two concurrent SPDC processes,
\mbox{$H_p\rightarrow H_s+V_i$}, and \mbox{$H_p\rightarrow
V_s+H_i$}, where $p$, $s$, $i$ represent the pump, signal, idler
fields, respectively, with $H$ ($V$) denoting the horizontal
(vertical) polarization.

\begin{figure}[tbh]
\centering
\includegraphics[width=0.45\textwidth]{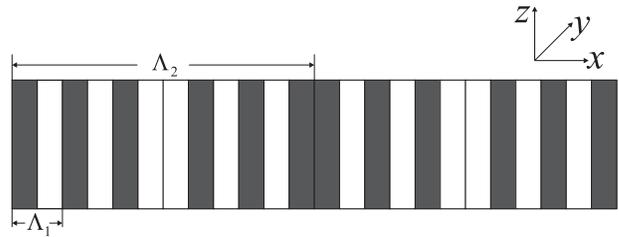}
\caption{Schematic of a dual-periodically poled potassium titanyl
phosphate crystal. Gray and blank areas are inverted ($-\chi^{(2)}$)
and background positive ($\chi^{(2)}$) domains, respectively.
}\label{fig:DPPKTP}
\end{figure}

The schematic of a dual-periodically poled KTP (DPPKTP) crystal is
shown in Fig.~\ref{fig:DPPKTP}, in which inverted domains (with
$-\chi^{(2)}$) distribute on a $+\chi^{(2)}$ background as a
dual-periodic structure. It is formed by twice-periodic modulation
of $\chi^{(2)}$. Suppose $g_1(x)$ and $g_2(x)$ are two periodic
functions as the sign of nonlinearity $\chi^{(2)}$. Then their
Fourier expansions can be written as
\begin{align}
  g_1(x)&=\sum_m\mathcal{G}_me^{-iG_mx},\\
  g_2(x)&=\sum_n\mathcal{G}_ne^{-iG_nx},
\end{align}
respectively, where the reciprocals are
\begin{equation}
  G_m=\frac{2m\pi}{\Lambda_1},\ \ \ \ G_n=\frac{2n\pi}{\Lambda_2},
\end{equation}
and the Fourier coefficients
\begin{equation}
  \mathcal{G}_m=\frac{2}{m\pi}\sin(mD_1\pi),\ \ \ \
\mathcal{G}_n=\frac{2}{n\pi}\sin(nD_2\pi).
\end{equation}
with $\Lambda_1$ and $\Lambda_2$ ($\Lambda_1<\Lambda_2$) denoting
the two modulation periods, $D_1$ and $D_2$  representing the duty
cycles, and nonzero integers $m$ and $n$ indicating the orders of
reciprocals. Then we can write the dual-periodic structure as
\begin{align}
  g(x)=g_1(x)g_2(x)=\sum_{m,n}\mathcal{G}_{m,n}e^{-iG_{m,n}x},
\end{align}
where
\begin{align}
  \label{eq:gmn1}\mathcal{G}_{m,n}=&\mathcal{G}_m\mathcal{G}_n=\frac{4}{mn\pi^2}\sin(mD_1\pi)\sin(nD_2\pi),\\
  \label{eq:gmn2}G_{m,n}=&G_m+G_n=\frac{2m\pi}{\Lambda_1}+\frac{2n\pi}{\Lambda_2}.
\end{align}
Then the modulation of the second-order nonlinear susceptibility
$\chi^{(2)}$ can be described as
\begin{equation}\label{eq:chi}
  \chi^{(2)}(x)=dg(x)=d\sum_{m,n}\mathcal{G}_{m,n}e^{-iG_{m,n}x},
\end{equation}
where $d$ is the effective nonlinear coefficient.

An arbitrary twice-periodic modulation could result in smaller
domains which may make fabrication more difficult. A straightforward
way to avoid the unwanted small domains, is by designing the
structure such that
\begin{equation}\label{eq:Dcondition}
\begin{cases}\Lambda_2/\Lambda_1=l/2\\
D_1=1/2\\
D_2=\lfloor l/2\rfloor/l
\end{cases},\ \ l\text{ is an integer bigger than }2,
\end{equation}
where $\lfloor\cdot\rfloor$ is the floor function to get the integer
part of a number. In practice, this condition can be satisfied by
tuning the temperature and wavelengths.

We consider the pump wave vector along the $x$ direction and $H$
($V$) in the $y$ ($z$) directions. By choosing the right wavelengths
and temperature we are able to obtain the following QPM conditions
for two backward-wave type SPDC processes
\begin{align}
  \label{eq:QPM1}\Delta k_1&=k_{p,H}-k_{s,H}+k_{i,V}-G_{m_1,n_1}=0,\\
  \label{eq:QPM2}\Delta k_2&=k_{p,H}-k_{s,V}+k_{i,H}-G_{m_2,n_2}=0,
\end{align}
where $G_{m_1,n_1}$ and $G_{m_2,n_2}$ are given by
Eq.~(\ref{eq:gmn2}) in the case of $\{m,n\}=\{m_1,n_1\}$ and
$\{m,n\}=\{m_2,n_2\}$, respectively. Here we require the two SPDC
processes to have the same signal frequency $\omega_s$ and the same
idler frequency $\omega_i$, with the energy conservation condition
$\omega_p=\omega_s+\omega_i$, where $\omega_p$ is the pump
frequency. In addition, as we shall see in Sec.~\ref{sec-SPDC}, we
require $m_1n_1=\pm m_2n_2$. In order to show the experimental
feasibility of such a structure, in the following we design two
possible structures based on the temperature-dependent Sellmeier
equation given by Emanueli and Arie~\cite{PPKTP:Emanueli2003}.

We first design a structure for a degenerate source of
$\lambda_p=655$ nm, $\lambda_s=\lambda_i=1310$ nm. Such a source
could find applications in long-distance fiber-based quantum
information processing, as the wavelength of the photons is in the
the second telecom window. At a working temperature of
$75^{\circ}$C, we get the two reciprocals for QPM as $G_{3,1}=17.47\
\mu\text{m}^{-1}$ and $G_{3,-1}=18.24\ \mu\text{m}^{-1}$,
corresponding to the two modulation periods $\Lambda_1=1.056\ \mu$m
and $\Lambda_2=16.36\ \mu$m, respectively. The ratio of the two
periods is $\Lambda_2/\Lambda_1=15.5$, and thus the duty cycle $D_2$
should be $15/31$.

We design a second structure for a non-degenerate source of
$\lambda_p=532$ nm, $\lambda_s=807.3$ nm, $\lambda_i=1560$ nm. This
choice is motivated by the photon source requirements in real-world
quantum networks, for example photonic memories in quantum
repeaters. The shorter-wavelength photon of this source can be used
for coupling and entangling atomic systems, and the other photon at
$1560$ nm can be transmitted over a long distance in fiber because
its wavelength lies in the low-loss transmission window of optical
fibers. By choosing the working temperature as $75.5^{\circ}$C,
we obtain the two reciprocals for QPM as $G_{3,1}=14.95\
\mu\text{m}^{-1}$ and $G_{3,-1}=15.93\ \mu\text{m}^{-1}$, with the
two corresponding modulation periods as $\Lambda_1=1.220\ \mu$m and
$\Lambda_2=12.82\ \mu$m, respectively, the ratio of which is
$\Lambda_2/\Lambda_1=10.5$, and therefore the duty cycle
$D_2=10/21$.

The above two example structures are both within current micron and
submicron periodic poling
techniques~\cite{PPKTP:Canalias2003,PPKTP:Canalias2005,Source:PP:backward:Canalias2007}.
In the following section we shall present the SPDC process in the
DPPKTP crystal and study the performances of the two example
sources.

\section{Generation of polarization-entangled photons}\label{sec-SPDC}

We consider a classical pump wave illuminating the DPPKTP crystal
with a length of $L$ in the $x$ direction and the interaction volume
denoted by $V$. The induced second-order nonlinear polarization is
given by~\cite{book:Boyd}
\begin{equation}
  P_i^{(2)}(\vec{r},t)=\varepsilon_0\chi_{ijk}^{(2)}E_j(\vec{r},t)E_k(\vec{r},t),
\end{equation}
where $\varepsilon_0$ is the vacuum dielectric constant and
$\chi_{ijk}^{(2)}$ is the second-order nonlinear susceptibility
tensor, where $i$, $j$, $k$ refer to the cartesian components of the
fields. Here, we use the Einstein notation of repeated indices for
tensor products. The Hamiltonian of the electromagnetic system can
be expressed as
\begin{equation}
  H=\frac{1}{2}\int_Vd^3\vec{r}\left(\vec{D}\cdot\vec{E}+\frac{1}{\mu_0}\vec{B}\cdot\vec{B}\right),
\end{equation}
where $\mu_0$ is the vacuum permeability constant. Since
$\vec{D}=\varepsilon_0\vec{E}+\vec{P}$, we obtain the interaction
Hamiltonian in the parametric down-conversion process
\begin{align}
  H_I(t)&=\frac{1}{2}\int_Vd^3\vec{r}\vec{P}\cdot\vec{E}\nonumber\\
&=\varepsilon_0\int_Vd^3\vec{r}\chi^{(2)}E_p(\vec{r},t)E_s(\vec{r},t)E_i(\vec{r},t),
\end{align}
where we replace $\chi_{ijk}^{(2)}/2$ with the second-order
nonlinear susceptibility $\chi^{(2)}$~\cite{book:Boyd}, which has
the form of Eq.~(\ref{eq:chi}) for an ideal structure. After
quantization of the electromagnetic fields, $E(\vec{r},t)$ becomes a
Hilbert space operator $\hat{E}(\vec{r},t)$, which can be decomposed
into its positive and negative parts
$\hat{E}(\vec{r},t)=\hat{E}^{(+)}(\vec{r},t)+\hat{E}^{(-)}(\vec{r},t)$.
Then we can rewrite the interaction Hamiltonianas
\begin{align}
  \label{eq:H1}
  \hat{H}_I(t)=&\varepsilon_0\int_Vd^3\vec{r}\chi^{(2)}(x)\hat{E}_p^{(+)}(\vec{r},t)\hat{E}_s^{(-)}(\vec{r},t)\hat{E}_i^{(-)}(\vec{r},t)\nonumber\\
   &+\text{H.c.},
\end{align}
where H.c. denotes the Hermitian conjugate part. Here, we only write
the two terms that lead to energy conserving processes, and we
neglect the other six terms that do not satisfy energy conservation
and are therefore of no importance in the steady state. Note that
neglecting these contributions is equivalent to making the
rotating-wave approximation.

Since the transverse structure of DPPKTP is homogeneous, we ignore
the transverse vectors of interacting waves and only consider the
interaction along the propagating direction. We consider the case of
signal and idler photons in forward and backward directions,
respectively. Then the negative parts of the field operators of the
signal and idler $\hat{E}_s$, $\hat{E}_i$ are represented by Fourier
integrals as
\begin{align}
   \label{eq:Es}\hat{E}_s^{(-)}(x,t)=&\sum_{q=H,V}\int
   d\omega_sE_{s,q}^{\ast}e^{-i(k_{s,q}x-\omega_st)}\hat{a}_{s,q}^{\dagger}(\omega_s),\\
   \label{eq:Ei}\hat{E}_i^{(-)}(x,t)=&\sum_{q=H,V}\int
   d\omega_iE_{i,q}^{\ast}e^{i(k_{i,q}x+\omega_it)}\hat{a}_{i,q}^{\dagger}(\omega_i),
\end{align}
where
$E_{j,q}=i\sqrt{\hbar\omega_j/(4\pi\varepsilon_0cn_q(\omega_j))}$,
$j=s,i$. For simplicity, here we consider a continuous-wave (cw)
plane-wave pump with horizontal polarization. In addition, the pump
field is treated as an undepleted classical wave, and thus the
positive part of its field operator is replaced with its complex
amplitude
\begin{equation}
 \label{eq:Ep}
  E_p^{(+)}(x,t)=E_pe^{i(k_{p,H}x-\omega_pt)}.
\end{equation}

Then by substituting Eqs.~(\ref{eq:chi}), (\ref{eq:Es}),
(\ref{eq:Ei}), and~(\ref{eq:Ep}) into Eq.~(\ref{eq:H1}), we obtain
\begin{align}
  \hat{H}_I(t)&=-\frac{\hbar E_P}{4\pi c}\sum_{q=H,V}\sum_{q'=H,V}\sum_{m,n}d\mathcal{G}_{m,n}
\int_{-L}^0 dx\int d\omega_s\nonumber\\
   &\times \int
   d\omega_i\sqrt{\frac{\omega_s\omega_i}{n_q(\omega_s)n_{q'}{(\omega_i)}}}\hat{a}_{s,q}^{\dagger}(\omega_s)\hat{a}_{i,q'}^{\dagger}(\omega_i)\nonumber\\
   &\times e^{i(\omega_s+\omega_i-\omega_p)t}e^{-i(k_{s,q}-k_{i,q'}-k_{p,H}+G_{m,n})x}+\text{H.c.},\label{eq:H2}
\end{align}

For the SPDC process, the interaction is weak, so under first-order
perturbation theory the state evolution from time $t'$ to $t$ can be
written as
\begin{equation}
  \ket{\Psi}=\ket{\text{vac}}+\frac{1}{i\hbar}\int_{t'}^t\hat{H}_I(\tau)d\tau\ket{\text{vac}}.\label{eq:SPDC1}
\end{equation}
Considering steady state output we may set $t'=-\infty$ and
$t=\infty$. Then we have
\begin{equation}
  \int_{-\infty}^{\infty}d\tau
  e^{i(\omega_s+\omega_i-\omega_p)\tau}=2\pi\delta(\omega_s+\omega_i-\omega_p),
\end{equation}
which gives the energy conservation relation
\begin{equation}\label{eq:energy}
  \omega_s+\omega_i-\omega_p=0.
\end{equation}
The integral over crystal length can be calculated as
\begin{align}\label{eq:PM}
  \int_{-L}^0dxe^{-i(k_{s,q}-k_{i,q'}-k_{p,H}+G_{m,n})x}=Lh(L\Delta k_{qq'}),
\end{align}
where $\Delta k_{qq'}=k_{p,H}-k_{s,q}+k_{i,q'}-G_{m,n}$ and the
$h$-function has the following form
\begin{equation}
 \label{eq:hx} h(x)=\frac{1-e^{-ix}}{ix}=e^{-i\frac{x}{2}}\text{sinc}\frac{x}{2}.
\end{equation}
$h(L\Delta k_{qq'})$ determines the natural bandwidth of the
two-photon state, as we shall see. In the case of infinite crystal
length, Eq.~(\ref{eq:PM}) becomes a $\delta$-function, thus leading
to the momentum conservation, i.e., the perfect phase-matching
condition, $\Delta k_{qq'}=0$.

Suppose that perfect phase matching conditions given by
Eqs.~(\ref{eq:QPM1}) and (\ref{eq:QPM2}) can be satisfied at
frequencies $\Omega_s$ and $\Omega_i$, with corresponding wave
vectors $K_{s,H}$, $K_{s,V}$, $K_{i,H}$, and $K_{i,V}$, such that
\begin{align}
  \Omega_s+\Omega_i=\omega_p,\ \ \ \ \ \
  K_{j,q}=\frac{n_q(\Omega_j)\Omega_j}{c},
\end{align}
with $j=s,i$ and $q=H,V$. Due to the existence of bandwidth, and
constrained by Eq.~(\ref{eq:energy}), we let
\begin{equation}
  \omega_s=\Omega_s+\nu, \ \ \ \ \ \ \omega_i=\Omega_i-\nu,
\end{equation}
where $|\nu|\ll\Omega_j$, $j=s,i$. Then in the case of the QPM
conditions given by Eqs.~(\ref{eq:QPM1}) and (\ref{eq:QPM2}), we can
write the state of SPDC as
\begin{align}\label{eq:SPDC2}
  \ket{\Psi}=&\ket{\text{vac}}+A_{HV}d_{HV}L\int
   d\nu h(L\Delta
k_{HV})\hat{a}_{s,H}^{\dagger}(\Omega_s+\nu)\nonumber\\
&\times\hat{a}_{i,V}^{\dagger}(\Omega_i-\nu)\ket{\text{vac}}+A_{VH}d_{VH}L\int
   d\nu h(L\Delta
k_{VH})\nonumber\\
&\times\hat{a}_{s,V}^{\dagger}(\Omega_s+\nu)\hat{a}_{i,H}^{\dagger}(\Omega_i-\nu)\ket{\text{vac}},
\end{align}
where
 \begin{align}
\label{eq:dhv} d_{HV}&=d\mathcal{G}_{m_1,n_1}=\frac{4d}{\pi^2m_1n_1}\sin\frac{m_1\pi}{2}\sin(n_1D_2\pi),\\
\label{eq:dvh} d_{VH}&=d\mathcal{G}_{m_2,n_2}=\frac{4d}{\pi^2m_2n_2}\sin\frac{m_2\pi}{2}\sin(n_2D_2\pi),\\
\label{eq:Ahv} A_{HV}&=\frac{iE_p}{2c}\sqrt{\frac{\Omega_s\Omega_i}{n_{s,H}n_{i,V}}},\\
\label{eq:Avh}
A_{VH}&=\frac{iE_p}{2c}\sqrt{\frac{\Omega_s\Omega_i}{n_{s,V}n_{i,H}}},
\end{align}
with $n_{j,q}$ denoting the refraction index of a photon with
polarization $q$ at frequency $\Omega_j$. Here $A_{HV}d_{HV}$ and
$A_{VH}d_{VH}$ are slowly varying functions of frequency, which have
been taken outside the integral.

We can see that the maximally polarization-entangled state can be
obtained under the condition of $|A_{HV}d_{HV}h(L\Delta
k_{HV})|=|A_{VH}d_{VH}h(L\Delta k_{VH})|$. The condition of
$d_{HV}=d_{VH}=d'$ can be satisfied straightforwardly by choosing
$m_1n_1=\pm m_2n_2$. In the following, we make calculations on
$h(L\Delta k_{HV})$ and $h(L\Delta k_{VH})$, i.e., the spectrum of
the photon pairs. In other words, the two-photon correlation time is
on the order of several hundred picoseconds.

\subsection{Characterizing the spectrum of photon pairs generated from our source}

We first expand the magnitudes of the wave vectors for signal and
idler photons around the central frequencies $\Omega_s$ and
$\Omega_i$ respectively, up to first order in $\nu$,
\begin{align}
  k_{s,q}&=\frac{n_q(\omega_s)\omega_s}{c}\approx
K_{s,q}+\frac{\nu}{u_q(\Omega_s)},\\
k_{i,q}&=\frac{n_q(\omega_i)\omega_i}{c}\approx
K_{i,q}-\frac{\nu}{u_q(\Omega_i)},
\end{align}
where $u_{q}(\Omega_j)=d\Omega_j/dK_{j,q}$ are the group velocities
of signal and idler photons at central frequencies, with $j=s,i$ and
$q=H,V$. Therefore we obtain
\begin{align}
 \label{eq:khv} \Delta
k_{HV}&=-\nu S_{HV},\ \ S_{HV}=\left[\frac{1}{u_H(\Omega_s)}+\frac{1}{u_V(\Omega_i)}\right],\\
\label{eq:kvh} \Delta k_{VH}&=-\nu S_{VH},\ \
S_{VH}=\left[\frac{1}{u_V(\Omega_s)}+\frac{1}{u_H(\Omega_i)}\right].
\end{align}

We thus obtain the joint spectral densities for the components
$\ket{H,V}$ and $\ket{V,H}$,
\begin{align}
 \label{eq:spectrum1} |h(L\Delta
k_{HV})|^2=&\text{sinc}^2\frac{\nu LS_{HV}}{2},\\
 \label{eq:spectrum2} |h(L\Delta
k_{VH})|^2=&\text{sinc}^2\frac{\nu LS_{VH}}{2},
\end{align}
and the corresponding bandwidths are
$\Delta\omega_{HV}\approx1.77\pi/(LS_{HV})$ and
$\Delta\omega_{VH}\approx1.77\pi/(LS_{VH})$, respectively. Compared
with the usual forward-wave type-II SPDC under the same conditions
on crystal length and frequencies~\cite{spectrum:rubin1994}, the
backward-wave source has a much narrower bandwidth, with a reducing
factor of $(u_H^{-1}+u_V^{-1})/|u_H^{-1}-u_V^{-1}|$.

More explicitly, we consider the two example structures given in
Sec.~\ref{sec-crystal}, and the crystal length is set to $2$ cm. For
the degenerate source, we get the bandwidth
$\Delta\omega_{HV}=\Delta\omega_{VH}\approx2\pi\times3.66$ GHz and
the reducing factor is $41$. For the non-degenerate source, we
obtain the two bandwidths as
$\Delta\omega_{HV}\approx2\pi\times3.61$ GHz and
$\Delta\omega_{VH}\approx2\pi\times3.63$ GHz, corresponding to
reducing factors of $25.9$ and $78.2$, respectively. Note that,
compared with the asymmetric spectrum in the forward-wave case, our
backward-wave source has an almost symmetric spectrum.

\subsection{Quantifying the polarization entanglement produced by our source}

To quantify the polarization entanglement produced by our source, we
employ a commonly used entanglement measure, namely,
concurrence~\cite{concurrence:Scott1997,*concurrence:def:Wootters1998},
whose value ranges from zero, for a non-entangled state, to one, for
a maximally entangled state. For a pure two-qubit state
$\ket{\psi}$, expressed in a fixed basis such as
$\{\ket{00},\ket{01},\ket{10},\ket{11}\}$, the concurrence
$C=|\bra{\psi}\sigma_y\otimes\sigma_y\ket{\psi}|$, where $\sigma_y$
is the second Pauli matrix $\left({0\atop i} {-i\atop 0}\right)$ in
the same basis. For our source, we need to treat the two-photon term
of the state given by Eq.~(\ref{eq:SPDC2}), denoted as
$\ket{\Psi_2}$. Note that the state $\ket{\Psi_2}$ is unnormalized
and the reciprocal of the square of its normalization constant is
the two-photon generation rate, given by
\begin{align}
  R=\langle\Psi_2\ket{\Psi_2}=&d'^2L^2\Big[|A_{HV}|^2\int
   d\nu\left|h(L\Delta
k_{HV}(\nu))\right|^2\nonumber\\
&+|A_{VH}|^2\int
   d\nu\left|h(L\Delta k_{VH}(\nu))\right|^2\Big].
\end{align}
Substituting Eqs.~(\ref{eq:spectrum1}) and (\ref{eq:spectrum2}) into
the above equation, we obtain
\begin{equation}\label{eq:p}
  R=2\pi d'^2L\left(\frac{|A_{HV}|^2}{S_{HV}}+\frac{|A_{VH}|^2}{S_{VH}}\right).
\end{equation}

Then we can calculate the concurrence
\begin{align}
  C=&\frac{|\bra{\Psi_2}\sigma_y\otimes\sigma_y\ket{\Psi_2}|}{\langle\Psi_2\ket{\Psi_2}}\nonumber\\
=&\frac{d'^2L^2|A_{HV}A_{VH}|}{R}\Big|\int
   d\nu h^{\ast}(L\Delta
k_{HV}(\nu))h(L\Delta k_{VH}(\nu))\nonumber\\
&+\int d\nu h^{\ast}(L\Delta k_{VH}(\nu))h(L\Delta
k_{HV}(\nu))\Big|.
\end{align}
By substituting Eqs. (\ref{eq:hx}), (\ref{eq:khv}) and
(\ref{eq:kvh}) into the above equation, we arrive at
\begin{align}
  C=\frac{2 S_{\text{min}}}{\delta_n S_{HV}+S_{VH}/\delta_n},
\end{align}
where $S_{\text{min}}=\min\{S_{HV},S_{VH}\}$ and
$\delta_n=\sqrt{n_{s,H}n_{i,V}/(n_{s,V}n_{i,H})}$.

For degenerate case, i.e., $\Omega_s=\Omega_i$, $S_{HV}=S_{VH}$,
$\delta_n=1$, and thus $C=1$,  so our source can generate degenerate
maximal polarization entanglement. This feature can also be seen
directly from the two-photon term of the state given by
Eq.~(\ref{eq:SPDC2}), which shows a maximally entangled state in the
form of $(\ket{HV}+\ket{VH})/\sqrt{2}$.

While for non-degenerate case, i.e., $\Omega_s\neq\Omega_i$,
$S_{HV}\neq S_{VH}$, $\delta_n\neq1$, and therefore $C<1$, so the
entanglement is nonmaximal. However, actually there is not a big
difference between $\delta_nS_{HV}$and $S_{VH}/\delta_n$, so the
concurrence is very near to $1$. Explicitly, let us consider the
example structure given in Sec.~\ref{sec-crystal}, the concurrence
of the entanglement generated from which is found to be as high as
$0.9978$.

\subsection{Generation rate of the entangled photon pairs}

The photon pair generation rate can be estimated from
Eq.~(\ref{eq:p}), by substituting Eqs.~(\ref{eq:Ahv}),
(\ref{eq:Avh}), and $|E_p|^2=2P/(\varepsilon_0n_pcS)$ into it, where
$P$ denotes the pump power and $S$ represents the transverse area of
the pump beam. Therefore we obtain
\begin{align}
  R=\frac{\pi
d'^2LP\Omega_s\Omega_i}{\varepsilon_0n_pc^3S}\left(\frac{1}{n_{s,H}n_{i,V}S_{HV}}+\frac{1}{n_{s,V}n_{i,H}S_{VH}}\right).
\end{align}

Let us consider the two specific example sources, for which we set
$P=1$ mW, $S=0.01$ mm$^2$, and $L=2$ cm. The nonlinear coefficient
$d'$ is given by Eq.~(\ref{eq:dhv}), where $m_1=3$, $n_1=1$, and the
effective nonlinear coefficient $d$, stemming from $d_{24}$, is
found to be $3.9$ pm$/$V. Then we find the generation rate of the
degenerate source to be $421$ pairs$/$s, and thus we get the
spectral brightness as $2\pi R/\Delta \omega\approx115$ pairs$/$(s
GHz mW). The generation rate of the non-degenerate source is found
to be $554$ pairs$/$s, corresponding to the spectral brightness of
$154$ pairs$/$(s GHz mW). We have to emphasize that the experimental
value of the photon pair rate and the two-photon spectrum will
definitely be affected by the poling quality, such as the deviations
and fluctuation of the poling period and duty
cycle~\cite{QPM:tuning:Fejer1992}. However, the state of the art of
the poling technique can enable us to engineer a nearly idealized
poled structure.

\section{Conclusions}
\label{sec-con}

In conclusion, we have presented a scheme for building
polarization-entangled photon pair sources utilizing backward-wave
type SPDC processes in a dual-periodically poled crystal. Our scheme
does not rely on any state projection and can work in degenerate and
non-degenerate cases. The backward-wave type SPDC enables the
entangled photon pairs from our source to transmit in a beam-like
way, exhibiting more efficient photon collection and mode
overlapping. Furthermore, the backward-wave type SPDC has a much
narrower bandwidth than the usual forward-wave one. In addition, our
scheme employs two concurrent SPDC processes in a single crystal
rather than any interferometer, and therefore our source is compact
and stable. By proper engineering on the domain structure a complete
set of Bell states can be achieved directly from this DPPKTP
crystal~\cite{source:PP:entangled:GuilletdeChatellus2006}. This
implies further applications in integrated photonic quantum
technologies.

We have designed two possible DPPKTP structures for degenerate and
non-degenerate sources, respectively. Using a $2$-cm-long bulk
crystal, the bandwidths of the two sources were found to be
$\sim3.6$ GHz with spectral brightnesses of $115$ and $154$
pairs$/$(s GHz mW), respectively. Our high-spectral-brightness
narrow-band sources should find applications in large-scale quantum
networks and other fields requiring narrow-band entangled photons.
Furthermore, we have also quantified the polarization entanglement
via concurrence and found that the degenerate source can provide
maximally polarization-entangled photon pairs while the concurrence
of the polarization entanglement generated from the non-degenerate
source is as high as $0.9978$. Finally, the two structures are both
within current manufacture technologies, and thus we believe our
sources can be realized in experiment. We hope our approach can
stimulate more investigations on applications of QPM on photonic
quantum technologies.

\begin{acknowledgments}
Y.X.G. thanks Bao-Sen Shi and Xu-Bo Zou for stimulating discussions.
This work was supported by the National Natural Science Foundation
of China (Grants No. 11004096, No. 10904066, No. 11004030, No.
11004029 and No. 11174052), and the State Key Program for Basic
Research of China (Grants No. 2011CBA00205 and No. 2012CB921802).
\end{acknowledgments}

\bibliography{E:/study/latex/bib/ref}
\end{document}